\newif\ifconfver
\confvertrue        

\ifconfver
\documentclass[10pt,twocolumn,twoside]{IEEEtran}
\else
\documentclass[11pt,draftcls,onecolumn]{IEEEtran}
\fi

\usepackage{tikz}
\usepackage{quantikz}
\usepackage{makecell}
\usepackage{bbding}
\usepackage{color,graphicx}
\usepackage{float}
\usepackage{multirow,amsmath,epsfig,amsfonts,amssymb,psfig,graphics,psfrag,theorem,calc,url,bm,cite}
\usepackage{stfloats,hyperref}
\usepackage{algorithm,algorithmic}
\usepackage{diagbox} 
\usepackage{hhline}
\usepackage{subfigure}
\usepackage{comment}
\newtheorem{theorem}{Theorem}
\usepackage{caption}
\usepackage{threeparttable}
\usepackage{url}
\usepackage{xcolor}
\usepackage{fancyhdr}
\usepackage{mathtools}

\makeatletter
\def\multilimits@{\bgroup
	\Let@
	\restore@math@cr
	\default@tag
	\baselineskip\fontdimen10 \scriptfont\tw@
	\advance\baselineskip\fontdimen12 \scriptfont\tw@
	\lineskip\thr@@\fontdimen8 \scriptfont\thr@@
	\lineskiplimit\lineskip
	\vbox\bgroup\ialign\bgroup\hfil$\m@th\scriptstyle{##}$\hfil\crcr}
\def\Sb{_\multilimits@}
\def\endSb{\crcr\egroup\egroup\egroup}
\makeatother

{\begin{list}{}%
		{\setlength{\rightmargin}{0pc}%
			\setlength{\leftmargin}{1pc} }} %
	{\end{list}}

\newlength{\twidth}
\ifconfver
\else
\fi

\definecolor{orange}{RGB}{255,107,0}

\theorembodyfont{\rmfamily}

{\begin{list}{}{
			\settowidth{\labelwidth}{\mbox{\textnormal{#1}}}%
			\setlength{\leftmargin}{\labelwidth+\labelsep}}}%
	{\end{list}}

\newcommand\bF{\ensuremath{{\bm F}}}

\newcommand\bI{\ensuremath{{\bm I}}}

\newcommand\bQ{\ensuremath{{\bm Q}}}

\newcommand\bT{\ensuremath{{\bm T}}}

\newcommand\bX{\ensuremath{{\bm X}}}
\newcommand\bY{\ensuremath{{\bm Y}}}

\definecolor{orange}{RGB}{255,107,0}

\ifconfver
\author{Chia-Hsiang Lin,~\IEEEmembership{Senior Member,~IEEE}, Po-Wei Tang,~\IEEEmembership{Student Member,~IEEE}, and Alfredo R. Huete}
\title{Quantum Feature-Empowered Deep Classification for Fast Mangrove Mapping
\thanks{This study was supported in part
by the Emerging Young Scholar Program (namely, the 2030 Cross-Generation Young Scholars Program) of National Science and Technology Council (NSTC), Taiwan, under Grant NSTC 113-2628-E-006-003; and in part by the Ph.D. Students Study Abroad Program of NSTC under Grant NSTC 113-2917-I-006-011.
We thank the National Center for Theoretical Sciences (NCTS) and the National Center for High-performance Computing (NCHC) for providing the computing resources.}
    \thanks{\textit{(Corresponding author: Chia-Hsiang Lin.)}}
    \thanks{Chia-Hsiang Lin. Lin is with the Department of Electrical Engineering, and with the Miin Wu School of Computing, National Cheng Kung University, Tainan 70101, Taiwan (R.O.C.) 
(e-mail: chiahsiang.steven.lin@gmail.com).}
     \thanks{Po-Wei Tang is with the Institute of Computer and Communication Engineering, Department of Electrical Engineering, National Cheng Kung University, Tainan 70101, Taiwan (R.O.C.) 
		(e-mail:  q38091526@gs.ncku.edu.tw).}
\thanks{Alfredo R. Huete is with the School of Life Sciences, University of Technology Sydney, Sydney, NSW 2007, Australia 
(e-mail: Alfredo.Huete@uts.edu.au).}	
}

\fancypagestyle{IEEEtitlepagestyle}{
	\fancyhf{} 
	\fancyhead[LE,RO]{\thepage} 
	\fancyfoot[C]{\scriptsize © 2024 IEEE. Personal use of this material is permitted. Permission from IEEE must be obtained for all other uses, in any current or future media, including reprinting/republishing this material for advertising or promotional purposes, creating new collective works, for resale or redistribution to servers or lists, or reuse of any copyrighted component of this work in other works. This is the accepted version of the manuscript. The final, published version is available at IEEE Xplore via https://doi.org/10.1109/TGRS.2024.3517459.}
	
}

\else

\fi
\begin{document}
	
	\bibliographystyle{IEEEtran}
	\maketitle
	\ifconfver \else \vspace{-0.5cm}\fi

\begin{abstract}
A mangrove mapping (MM) algorithm is an essential classification tool for environmental monitoring.
The recent literature shows that, compared to other index-based MM methods that treat pixels as spatially independent, convolutional neural networks (CNNs) are crucial for leveraging spatial continuity information, leading to improved classification performance.
In this work, we go a step further to show that quantum features provide radically new information for CNN to further upgrade the classification results.
Simply speaking, CNN computes affine-mapping features, while quantum neural network (QNN) offers unitary-computing features, thereby offering a fresh perspective in the final decision-making (classification).
To address the challenging MM problem, we design an entangled spatial-spectral quantum feature extraction module.
Notably, to ensure that the quantum features contribute genuinely novel information (unaffected by traditional CNN features), we design a separate network track consisting solely of quantum neurons with built-in interpretability.
The extracted pure quantum information is then fused with traditional feature information to jointly make the final decision.
The proposed quantum-empowered deep network (QEDNet) is very lightweight, so the improvement does come from the cooperation between CNN and QNN (rather than parameter augmentation).
Extensive experiments will be conducted to demonstrate the superiority of QEDNet.


\end{abstract}

	\ifconfver \else \vspace{-0.0cm}\fi
	
	\ifconfver \else \vspace{-0.5cm}\fi
	
	\ifconfver \else  \fi

\section{Introduction}\label{sec:intro}
Mangrove forests are known for their exceptional above-ground biomass and carbon storage capacity.
They are a critical part of blue carbon ecosystems (coastal systems) \cite{jones2020estimating,nellemann2009blue}, which are known for their ability to trap carbon, typically consisting of mangroves, seagrasses, and salt marshes.
{In addition to their role in carbon storage, mangrove forests are essential habitats for numerous species, greatly enhancing coastal biodiversity.
Furthermore, mangroves play a crucial role in protecting coastlines from erosion, aligning with the United Nations Sustainable Development Goals (SDGs) 13, 14, and 15 \cite{kazemi2021mangrove,Siikamaki14369,chow2018mangrove}.}
Remarkably, because of direct human impact and global climate change, mangrove habitats are rapidly declining.
Thus, the importance of studying the blue carbon ecosystem (i.e., monitoring mangrove distributions) has been rapidly increasing, aligning with long-term SDG achievements \cite{carugati2018impact,zeng2021global}.
However, obtaining the direct distribution of mangroves may be challenging due to their location in intertidal zones, which could impede on-site investigation because of muddy environments, tidal effects, and dense forests \cite{zhu2020estimating,chen2023mapping,zhang2021comparison}.

Hence, a more practical approach to investigating mangrove distribution would be using satellite images.
Satellite images capture a wide range of spectral wavelengths from visible to shortwave infrared regions (VSWIR), providing diverse spectral information that contributes to significant advancements \cite{BPCSNC2013,reviewaccess14,lin2021all,lin2024signal}.
Specifically, it has extensive applications in the remote sensing area, such as land cover classification, change detection, environmental monitoring, and mangrove mapping (MM) \cite{Hongtgrs2021,Houtgrs2022,codehcd,underwood2006mapping,codemm}.
Even though remote sensing images may encounter issues such as inpainting and dehazing problems, several newly developed algorithms are created to rectify corrupted data as a preprocessing step \cite{LinTNNLS2023,hyperqueen,tang2024transformer}.
On the other hand, compared to the time and human-resource-consuming geography field research, remote sensing data demonstrates its advantages in efficiently acquiring mangrove distribution due to its extensive spatial coverage and easier access to quality data.
With its strong material identifiability across various wavelengths \cite{kuenzer2011remote, green1998remote, lin2018maximum}, identifying the underlying mangrove regions from the observed land cover in satellite images is achievable.
Especially remote sensing data covers critical spectral bands highly correlated with mangrove properties, including the green band, near-infrared (NIR) band, and shortwave infrared (SWIR) bands \cite{drusch2012sentinel}.

Existing MM algorithms can broadly be classified as index model-based and neural network (NN) model-based methods \cite{MVI_paper,iovan2020deep}.
Specifically, the index model-based method utilizes the water/vegetation-related bands to estimate the mangrove distributions.
On the other hand, the NN model-based method learns the non-linear function $f_{\theta}(\cdot)$ by iteratively updating the network parameters $\theta$ through the back-propagation algorithm \cite{bp},  thereby generating the classification maps from the satellite imagery. 
Note that recent literature has examined the advantages and disadvantages of MM methods using various imagery types, such as synthetic aperture radar (SAR), light detection and ranging (LiDAR), and aerial imagery \cite{maurya2021remote}.

Next, we will sequentially introduce recently developed algorithms for MM.
Vegetation index (VI) methods typically utilize spectral characteristic information to obtain classification maps \cite{HUETE1988295,HUETE2002195}.
For example, Tucker \cite{tucker1979red} proposed the normalized difference vegetation index (NDVI) for monitoring vegetation by using red and infrared bands in the form of (NIR-Red)/(NIR+Red).
Diniz \textit{et al.} \cite{diniz2019brazilian} proposed a modular mangrove recognition index (MMRI) for discriminating Brazilian mangroves, which utilizes NDVI and the modified normalized difference water index (MNDWI) \cite{xu2006modification} in the form of 
($\mid$MNDWI$\mid$-$\mid$NDVI$\mid$)/($\mid$MNDWI$\mid$+$\mid$NDVI$\mid$).
Baloloy \textit{et al.} \cite{MVI_paper} proposed a mangrove vegetation index (MVI) for accurately mapping the mangrove extent, which jointly uses Sentinel-2 green, NIR, and SWIR bands in the form of (NIR-Green)/(SWIR1-Green).
In summary, MVI is designed to differentiate between the distinct greenness and moisture of mangroves from different landscapes.
As an advanced version, Yang \textit{et al.} \cite{YANG2022236} proposed an enhanced mangrove vegetation index (EMVI) based on hyperspectral images, aiming at enhancing the difference in greenness and canopy moisture content between mangroves and other vegetation using a green band and two shortwave-infrared bands in the form of (Green-SWIR2)/(SWIR1-Green), thereby successfully identifying mangrove distribution.
By further considering random forest techniques, Zhao \textit{et al.} \cite{ZHAO2023209} proposed an interpretable MM approach (IMMA), consisting of five features, including B12, B8/B2, mangrove vegetation index (MVI), normalized difference index (NDI), and elevation from a digital elevation model.

As NN rapidly developed, Iovan \textit{et al.} \cite{iovan2020deep} proposed a deep convolutional neural network (DCNN) with multiple dense layers to extract spatial features and classify the mangrove distribution, achieving satisfactory results on both WorldView-2 and Sentinel-2 imageries.
Different from DCNN, Dong \textit{et al.} \cite{dong2021gc} adopted global context blocks and the classical UNet framework (GC-UNet) \cite{ronneberger2015u} with short connection mechanisms to retain long-range dependency information.
Besides, low-level and high-level features were fused by adaptively spatial feature fusion blocks to enhance the performance of the proposed algorithm, which was evaluated on the Landsat 8 dataset.
From a different perspective, Guo \textit{et al.} \cite{capsulenetrs21} introduced the capsule concept into the basic UNet structure to form a Capsules-UNet, meanwhile addressing the problem of excessive memory burden and a large number of parameters.
Specifically, Capsules-UNet replaced max-pooling layers with convolutional strides and dynamic routing \cite{lalonde2018capsules} to produce more accurate MM results.
By further considering multiple inputs, Guo \textit{et al.} \cite{MENetrs21} proposed another UNet-based network, utilizing normalization difference vegetation index (NDVI), modified normalized difference water index (MNDWI), forest discrimination index (FDI), wetland forest index (WFI), mangrove discrimination index (MDI), and the first principal component analysis (PCA) feature.
Furthermore, the global attention module, multiscale context embedding, and boundary fitting unit are utilized for mangrove extraction, thus enhancing the accuracy of the classification process.

Not only considering NN technology, Lin \textit{et al.} \cite{codemm} intriguingly integrated convex optimization (CO) and deep learning (DE) to classify mangrove distributions (CODE-MM) using the novel CODE theory, which was initially developed to address the hyperspectral inpainting problem in scenarios where the training data is limited.
{Specifically, CODE-MM extracts features from a rough DE solution by using the $\bQ$-norm as a deep regularizer.
Therefore, benefiting from the CODE framework \cite{LinTGRS2021}, CODE-MM achieves satisfactory classification results even with small data.}


An MM algorithm is an important classification tool for environmental monitoring.
{Surprisingly, benchmark index-based methods rely solely on spectral characteristics and do not take spatial continuity into account.
However, recent research \cite{codemm} has highlighted the significance of convolutional neural networks (CNNs) in improving classification performance by effectively utilizing adjacent spatial information.
Besides, benchmark index-based methods also require users to manually tune the threshold parameter to obtain satisfactory classification results.
Thus, it is necessary to develop end-to-end CNN without manually tuned parameters for user-friendliness.
Not to mention, in complex scenarios, establishing an optimal threshold for index-based methods to distinguish mangroves from other land covers is challenging and time-consuming.
In contrast, NN-based methods trained with ground truth can automatically determine the parameters based on validation performance.
Remarkably, quantum deep learning technology has recently achieved outstanding performance in some challenging satellite remote sensing missions (e.g., image restoration) \cite{hyperqueen} and well-log interpretation tasks \cite{liu2021quantum}, showing the value of quantum features in technology advancements.}

Hence, this work aims to design an end-to-end quantum-empowered deep network (QEDNet) without handcrafted parameter tuning, where quantum-entangled features will be further integrated into CNN to upgrade the MM performances.
In summary, CNN focuses on affine mapping features, while quantum neural networks (QNN) provide unitary computing capabilities during feature extractions \cite{hyperqueen}, thus obtaining features from a radically new approach for better final decision/classification.
Additionally, ablation studies are conducted to assess how quantum features introduce new information for enhancing the classification performance, through trickily designed independent but symmetric CNN and QNN tracks.

Unlike index-based methods, the proposed QEDNet is an automatic algorithm that does not require threshold settings.
Furthermore, it delivers quick computational performance as both branches can be implemented efficiently.
QEDNet has made significant progress in mapping mangroves, as demonstrated by its exceptional results analyzed in Section \ref{sec: experiment}. 
The main contributions are summarized as follows:

\begin{itemize}
\item 
We propose the interpretable QEDNet, which consists of independent CNN and QNN branches.
QEDNet distills and fuses useful CNN and QNN features, thereby generating effective classification maps for various testing scenarios across different nations.

\vspace{0.1cm}
\item 
The QNN branch comprises spatial and spectral feature extractors and entangled feature fusion modules.
Specifically, the spatial feature extractor combines and entangles local spatial information, while the spectral one refines features along the spectral dimension.
Then, group-wise fusion captures the global channel correlation from local dependencies.
Moreover, shortcut connections are adopted to mitigate the barren plateau effect (a.k.a. quantum gradient vanishing effect) \cite{hyperqueen}.

\vspace{0.1cm}

\item 
As demonstrated in Section \ref{sec:ablation}, QEDNet is a simple yet effective framework.
Entangled unitary-computing QNN features are fused into the affine-mapping CNN features, thereby providing additional valuable information.
Furthermore, QEDNet is a lightweight network that provides faster computational time than other benchmark deep learning-based methods.
Though index-based methods are even faster, the spatial continuity is ignored; thus, their MM performances are not comparable to QEDNet. 
\end{itemize}

The remaining parts of this paper are organized as below.
In Section \ref{sec: method}, we will provide a detailed illustration of the proposed QEDNet.
In Section \ref{sec: experiment}, we present and analyze extensive experiments conducted in different countries. 
In particular, we conduct ablation studies to demonstrate the effectiveness of the proposed dual-branch framework, to assess the efficacy of the quantum features in Section \ref{sec:ablation}.
Finally, we summarize our conclusions and insights in Section \ref{sec: conclusion}.

\section{Method} \label{sec: method}
\begin{figure*}[t]
\begin{center}
\includegraphics[width=0.99\textwidth] {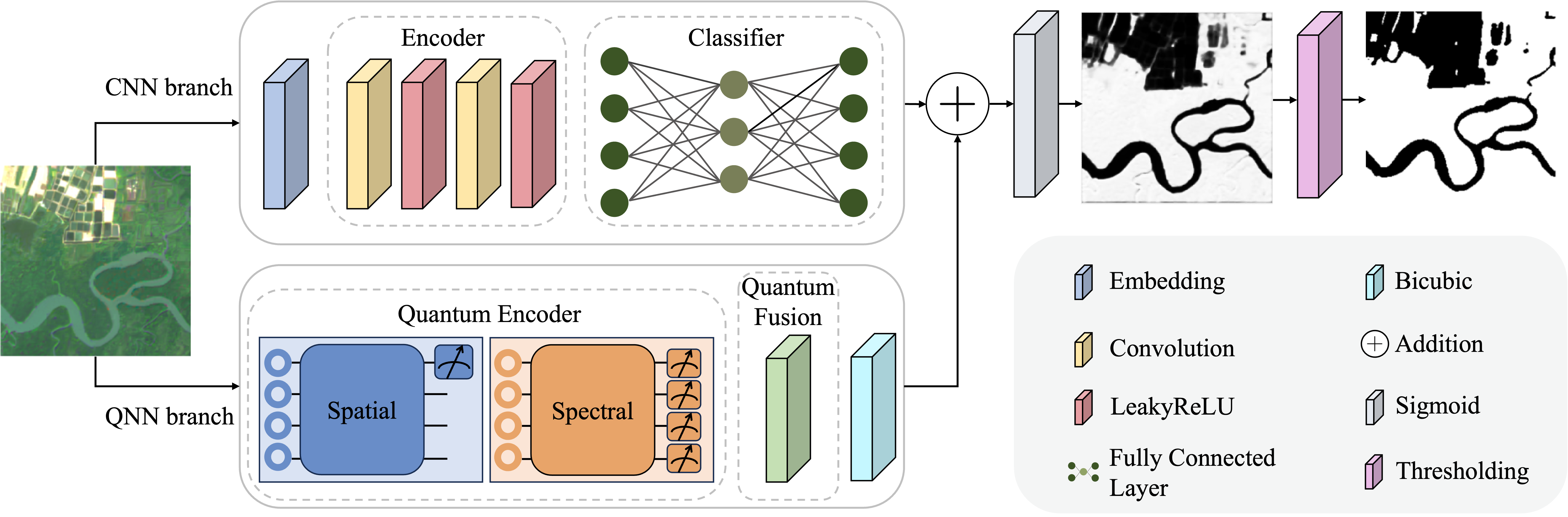}
\end{center}
\caption{Graphical illustration of the proposed quantum-empowered deep network (QEDNet).
It incorporates the entangled quantum features into traditional CNN under a parallel structure, so that both QNN and CNN features can equally contribute to the final classification results.
Precisely, QNN captures unitary-computing feature due to the nature of quantum neurons, and this is fundamentally different from the affine-mapping CNN feature, thus further providing new information for making better final decision/classification.
For the QNN branch, the detailed structure of the quantum spatial-spectral encoder will be presented in Figure \ref{fig:qnnencoder}, and the quantum fusion module will be presented in Figure \ref{fig:fusion}.}
\label{fig:framework}
\end{figure*}

\begin{figure*}[t]
\begin{center}
\includegraphics[width=1\textwidth] {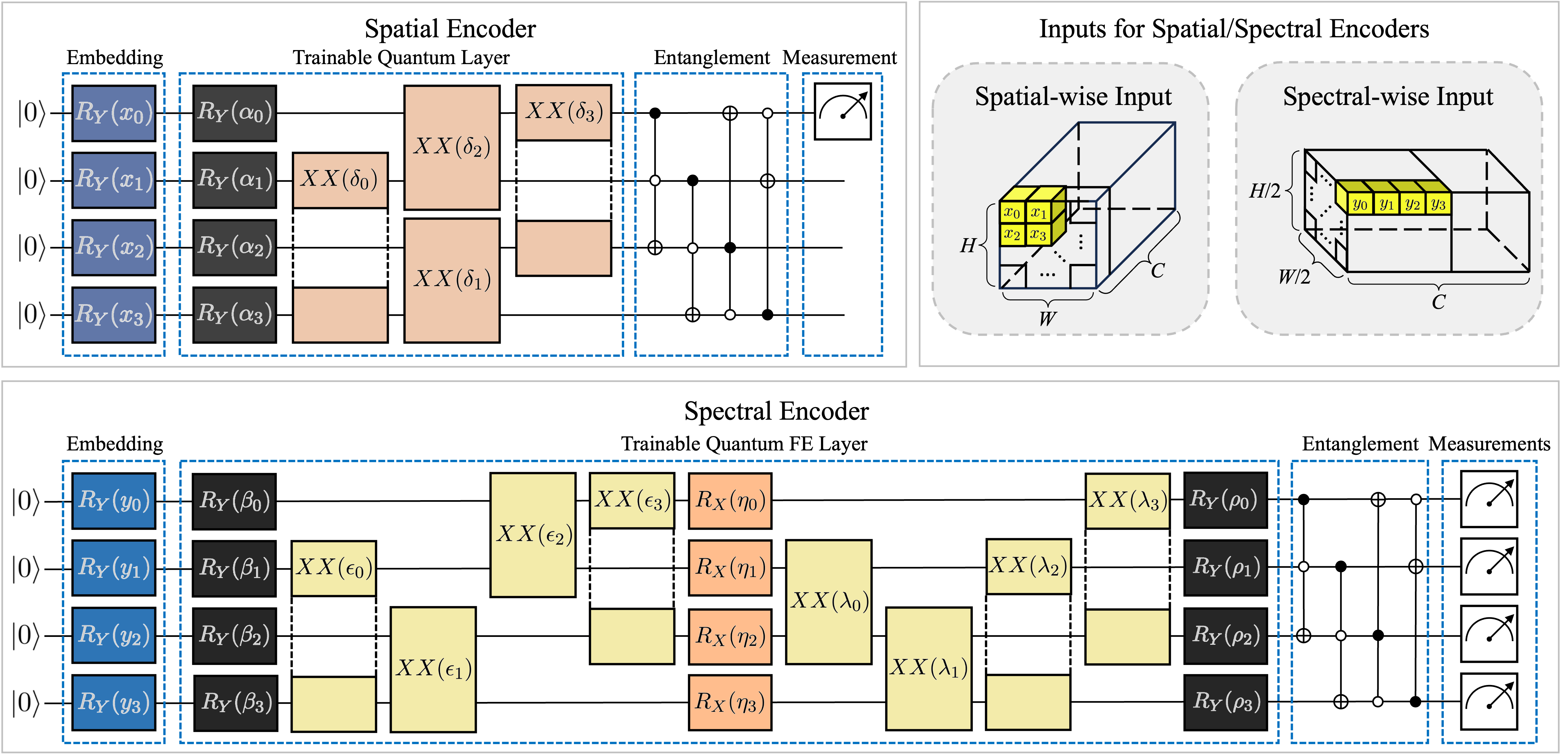}
\end{center}
\caption{Detailed architecture of the proposed QNN-based spatial encoder and spectral encoder, both composed of quantum unitary operators.
Specifically, the spatial encoder entangles/compresses spatial information from each $2\times 2$ local patch into a single pixel, where the local patch is illustrated as the yellow $x_i$'s serving as one input batch of the spatial encoder.
Also, the spectral encoder entangles/captures quantum characteristics across various channel groups (cf. Figure \ref{fig:fusion}) along the spectral dimension, where one channel group is illustrated as the  yellow $y_i$'s serving as one input batch of the spectral encoder.}
    \label{fig:qnnencoder}
\end{figure*}
\begin{figure}[t]
\begin{center}
\includegraphics[width=0.49\textwidth] {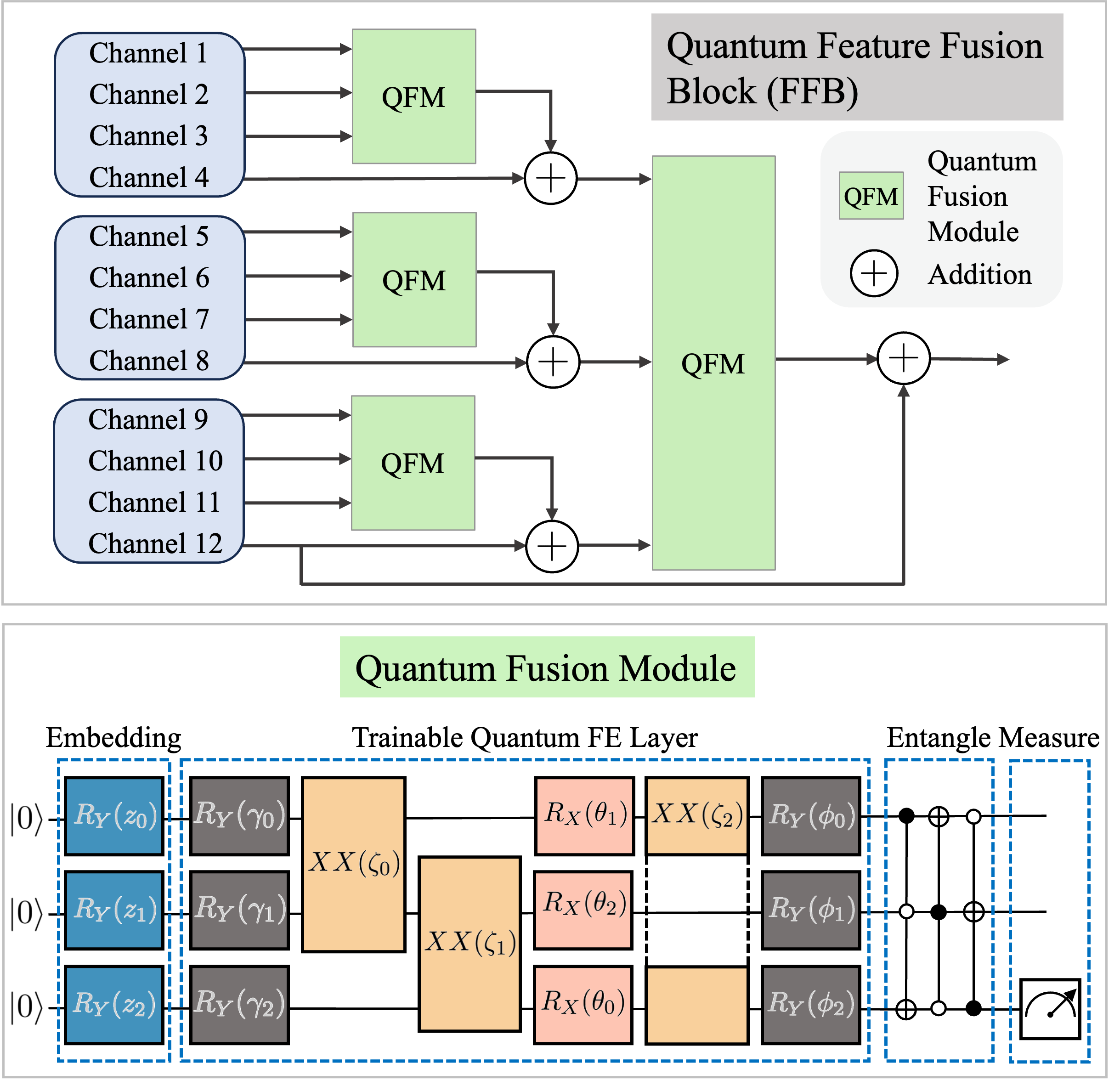}
\end{center}
\caption{Detailed architecture of the proposed QNN-based feature fusion block (FFB).
The quantum FFB aims to further integrate the previously encoded spatial-spectral quantum features while capturing the correlation between the neighboring bands.
To obtain the global channel relations in the final QNN features, we first set three local groups to capture their neighboring channel correlations using the quantum fusion module (QFM), and then fuse the three representative locally correlated features.
The QFM is designed by rotation gates and Ising gates, followed by entanglement mechanisms, where the used gates are defined in Table \ref{tab:common_qu_gate}.}
    \label{fig:fusion}
\end{figure}

Compared to index-based methods that ignore spatial continuity of the mangrove distribution, CNN has exhibited its effectiveness in upgrading classification results by extracting spatial features \cite{codemm}.
Motivated by this fact, we design an explainable quantum-empowered deep network (QEDNet) to tackle the challenging MM problem by combining quantum deep learning and CNN techniques. 
As illustrated in Figure \ref{fig:framework}, the proposed QEDNet cooperatively fuses the CNN and QNN features to determine the final classification results, for which two independent but symmetric CNN and QNN tracks are customized.
Remarkably, we integrate the lightweight CNN/QNN branches in a parallel structure without using intense layers, thereby demonstrating the effectiveness does come from the new information brought by the quantum computing (rather than the heavy network parameters).
Precisely, the QNN branch captures entangled unitary features \cite{hyperqueen}, fundamentally different from CNN's affine mapping features, thus providing radically new information for decision-making.

Therefore, the customized QNN branch, composed of a quantum-empowered spatial-spectral feature extractor and a quantum feature fusion block (FFB), is designed to extract entangled unitary-computing QNN features, as detailed in Figures \ref{fig:qnnencoder} and \ref{fig:fusion}.
Ablation study will also be conducted in Section \ref{sec:ablation} to show that the QNN track is indeed critical, and if it is replaced by an equally-capable CNN track, the MM results become weaker.
The design philosophy and physical meanings of the proposed QEDNet will be hierarchically presented in Sections \ref{sec:two-branch}, \ref{sec:CNN}, and \ref{sec:QNN}.

\subsection{Dual-Branch QEDNet for Quantum Feature and Traditional Feature Extractions} \label{sec:two-branch}

With the advancement of deep learning technology, more and more inputs/features are utilized to capture additional information from different perspectives to improve classification performance.
For instance, recent literature \cite{MENetrs21} has explored the utilization of multiple water/vegetation indices and the first principal component analysis (PCA) feature as the inputs for training the MM network.
Furthermore, inspired by dual-branch neural networks for studying different data modalities \cite{wang2018learning} with multiple features, it is natural to develop a multiple-branch network for solving the challenging MM problem.
Intuitively, if we extract additional features embedding new information from different perspectives, one can expect that it will lead to improved classification performance.
Therefore, we propose a radically new quantum feature extraction scheme (cf. quantum spatial-spectral encoder presented in Figure \ref{fig:qnnencoder} and the quantum fusion module presented in Figure \ref{fig:fusion}), and use it to obtain new quantum information for the classification task.

For better comprehension, Figure \ref{fig:framework} illustrates the core concept of the proposed dual-track framework, including the independent CNN branch and QNN branch, for extracting information from two fundamentally different perspectives.
%
%
To sum up, we will design an interpretable quantum-empowered deep network (QEDNet) by integrating CNN and QNN features to collaboratively achieve state-of-the-art MM performance.
Therefore, the overall architecture of the proposed QEDNet is defined as follows:
\begin{align}
\bY=\text{Sigmoid}(f_{\text{CNN}}(\bX)+f_{\text{QNN}}(\bX)),\label{eq:twobranch}
\end{align}
where $\bY\in \mathbb{R}^{H\times W}$ represents the classification result indicating the likelihood of being mangrove pixels; $f_{\text{CNN}}(\cdot)$ denotes the CNN branch function; 
$f_{\text{QNN}}(\cdot)$ denotes the QNN branch function; 
$\bX\in \mathbb{R}^{H\times W\times C}$ is the Sentinel-2 multispectral image (MSI) with $C=12$ bands/channels; 
$\text{Sigmoid}(\cdot)$ denotes the sigmoid function returning a value in $[0,1]$.
We remark that the QEDNet will be designed for Sentinel-2 satellite in this paper as it provides data for numerous popular MM methods \cite{MVI_paper,codemm}, but the theory and architecture proposed here are generally applicable to other multispectral satellite data.

Specifically, we employ a simple yet effective dual-track framework with CNN and QNN branches, rather than designing a single complex/deep CNN network to address the challenging MM problem.
The process combines useful CNN and QNN features for enhancing the network capability to identify mangrove pixels.
Benefiting from the symmetric lightweight framework, the proposed QEDNet results in faster computational speed than existing benchmark deep learning-based MM methods.
By Equation \eqref{eq:twobranch}, the dual-track model using informative CNN and QNN features can be elegantly implemented without relying on the weight attention mechanism, which requires additional learning procedure.
Finally, after removing ten percent of outliers, a threshold value is automatically set as half of the maximum sigmoid value \cite{codemm}, and is used to transform the sigmoid map into a binary classification map, as illustrated in Figure \ref{fig:framework}.
This automatic thresholding procedure is much more user-friendly especially when comparing to index-based methods that require handcrafted threshold tuning.

\subsection{CNN Branch} \label{sec:CNN}

First, one should notice that benchmark deep learning-based methods \cite{dong2021gc,capsulenetrs21,MENetrs21} mainly adopt the deeper scheme, such as basic UNet structure \cite{ronneberger2015u} for solving challenging MM problems.
Not to mention that these methods will further utilize encoding modules under the UNet baseline, making their model architecture very deep.
From very different perspectives, we develop a lightweight dual-branch model by combining CNN and QNN tracks in this work.
To echo the design philosophy, the CNN branch is developed as a lightweight model rather than a complex/deep network.
As shown in Figure \ref{fig:framework}, the CNN branch only contains three components: an embedding layer (i.e., $f_{\theta_\text{E}}(\cdot)$), an encoder block (i.e., $\text{Encoder}(\cdot)$), and fully connected layers (i.e., FCL$(\cdot)$).
{The embedding layer consists of a single $3\times3$ convolutional layer with padding set as 1.}
Specifically, the embedding layer helps transform the input from the image domain into the feature domain.
After the embedding stage, the encoder block, comprising $3 \times 3$ convolutional layers and the LeakyReLU function, is adopted to extract useful features.
Subsequently, fully connected layers (FCL) integrate previously encoded features by channel-wise interconnections, allowing the CNN branch to generate final representative CNN features as output.

As illustrated in Figure \ref{fig:framework}, the overall architecture of the proposed CNN branch $f_{\text{CNN}}(\cdot)$ can be defined as follows:
\begin{align}
\text{Encoder}(\bT)&=\sigma(f_{\theta_2}(\sigma(f_{\theta_1}(\bT)))),\\
f_{\text{CNN}}(\bT)&=\text{FCL}(\text{Encoder}(f_{\theta_\text{E}}(\bT))),
\end{align}
where $f_{\theta_i}(\cdot)$ represents the $i$th convolutional layer in the encoder block and $\theta_i$ denotes the corresponding learning weights in $f_{\theta_i}(\cdot)$; $\sigma(\cdot)$ is the LeakyReLU activation function with negative slope $0.2$ \cite{maas2013rectifier}; $\bT$ denotes the 3-D input tensor; $f_{\theta_\text{E}}(\cdot)$ represents the embedding layer and $\theta_\text{E}$ denotes the corresponding learning weights in $f_{\theta_\text{E}}(\cdot)$.
Finally, the CNN feature used in Equation \eqref{eq:twobranch} can be easily obtained as follows:
\begin{align}
\bF_{\text{CNN}}&=f_{\text{CNN}}(\bX),
\end{align}
where $\bF_{\text{CNN}}\in \mathbb{R}^{H\times W }$ represents the CNN feature extracted from the CNN branch for determining the final classification result.
By following the aforementioned architecture, the lightweight CNN branch can be effectively implemented.

\begin{table}[t]
\scriptsize
\setlength{\tabcolsep}{5.8pt} 
\caption{Symbols and mathematical definitions for the quantum gates used in the QNN branch $f_{\text {QNN}}(\cdot)$ (cf. Figure \ref{fig:qnnencoder} and Figure \ref{fig:fusion}), all corresponding to some unitary operators, where $\theta$ represents the learnable parameters.  
For conciseness, we use $\delta$ and $\gamma$ to denote $\cos(\frac{\theta}{2})$ and $\sin(\frac{\theta}{2})$, respectively.
Also, DIAG($A$, $B$, $C$) is a block-diagonal matrix with diagonal blocks $A$, $B$ and $C$, and $\bI_n$ denotes the $n\times n$ identity matrix.}\label{tab:common_qu_gate}
\begin{center}
\renewcommand\arraystretch{1.3}
\begin{tabular}{|c c c|} 
 \hline
 \rule{0pt}{2ex}
 Quantum Gate & Symbol & Unitary Operator 
 \rule{0pt}{2ex}
 \\
 \hline
 \rule{0pt}{4ex}
 Rotation X
 &
 \begin{tikzcd}
    \qw & \gate{R_{X}(\theta)} & \qw
 \end{tikzcd}
 &
 $\begin{pmatrix}
    \delta & -i \gamma \\
    -i \gamma & \delta
\end{pmatrix}$
 \rule{0pt}{4ex}
 \\ 
 \hline
 \rule{0pt}{4ex}
 Rotation Y
 &
 \begin{tikzcd}
    \qw & \gate{R_{Y}(\theta)} & \qw 
 \end{tikzcd}
 & 
 $\begin{pmatrix}
    \delta & - \gamma \\
    \gamma & \delta
\end{pmatrix}$
 \rule{0pt}{4ex}
\\
 \hline
 \rule{0pt}{4ex}
 \rule{0pt}{6.5ex}
 Ising XX
&
\begin{quantikz}
    \qw & \gate{XX(\theta)} & \qw
\end{quantikz}
&
$\begin{pmatrix}
    \delta & 0 & 0 & -i\gamma\\
    0 & \delta & -i\gamma & 0\\
    0 & -i\gamma & \delta & 0\\
    -i\gamma & 0 & 0 & \delta
\end{pmatrix}$
\\
\hline
Pauli-Z &
\begin{tikzcd}
\meter{} 
\end{tikzcd}
&
$\begin{pmatrix}
    1 & 0 \\
    0 & -1 \\
\end{pmatrix}$
 \\
 \hline
NOT &
\begin{tikzcd}
\qw & \gate{X} & \qw
\end{tikzcd}
&
$\begin{pmatrix}
    0 & 1 \\
    1 & 0 \\
\end{pmatrix}$
 \\
 \hline
 \rule{0pt}{11ex}
Toffoli (CCNOT)
 &
 \begin{tikzcd}
    \qw & \ctrl{1} & \qw \\
    \qw & \octrl{1} & \qw \\
    \qw & \targ{} & \qw
 \end{tikzcd}
 &
 $\textrm{DIAG}(\bm{I}_4,X,\bm{I}_2)$
 \rule[-5ex]{0pt}{4ex}
 \\
 \hline
\end{tabular}
\end{center}
\end{table}

\subsection{QNN Branch}\label{sec:QNN}
Unlike traditional CNN whose features basically come from some specific affine computing, the QNN provides novel feature information obtained from unitary computing with entanglement scheme \cite{hyperqueen}.
This inspires us to fuse both affine CNN feature and unitary QNN feature by developing the dual-branch network architecture (cf. Figure \ref{fig:framework}) for better classification results.
Precisely, the QNN track consists of a quantum spatial-spectral feature encoder and the quantum fusion module (QFM), as illustrated in Figure \ref{fig:qnnencoder} and Figure \ref{fig:fusion}. 
Before introducing the design philosophy and implementation details, the symbol and mathematical definition of each quantum neuron adopted in the QNN track are summarized in Table \ref{tab:common_qu_gate}.
Note that as the quantum neurons are all implemented via some Hamiltonian for a specific time, the Schrödinger equation implies that the operators corresponding to those quantum neurons are all unitary \cite{Qunitary}, as can be seen from Table \ref{tab:common_qu_gate}.
As shown in Figures \ref{fig:qnnencoder} and \ref{fig:fusion}, Rotation X, Rotation Y, Ising XX, Pauli-Z, and Toffoli quantum gates \cite{lin2024primeblindmultispectralunmixing} are adopted to build the explainable quantum feature extraction modules with explicit physical meanings, as introduced below.

First, we aim to extract features with spatial continuity, ensuring that the QNN feature contains useful spatial information \cite{zhu2020residual}.
The QNN-based spatial feature extractor operates on each $2 \times 2$ local patch, using the $R_Y$-Ising$^3$-Toffoli$^4$ architecture that empirically works very well, as illustrated in Figure \ref{fig:qnnencoder}.
Through this quantum encoding mechanism, it first extracts spatial information from four neighboring pixels (cf. yellow $x_i$'s in Figure \ref{fig:qnnencoder}) that jointly serve as one input batch of the spatial encoder.
Subsequently, the obtained spatially entangled features are fused into one single representative pixel, and its physical meaning can be considered as spatial fusion/compression.
{While this step may seem similar to kernel convolution and pooling operations in conventional CNNs, the quantum feature extractor operates fundamentally differently, relying on unitary computing, entanglement, and measurement.}
Hence, the quantum spatial encoder adaptively learns how to generate the fused/compressed QNN features.
Besides, the local $2\times2$ patches (cf. yellow $x_i$'s in Figure \ref{fig:qnnencoder}) can be considered as batches in the training stage, indicating that all the patches efficiently share network weights for extracting uniform local spatial features.
This strategy significantly reduces the network parameters, thereby leading to the desired lightweight network design in QEDNet.

Second, after obtaining the spatial feature from the quantum spatial encoder, we further extract its feature along the spectral dimension.
The QNN-based spectral feature extractor operates on each $1 \times 4$ channel-wise sequence, using the $R_Y$-Ising$^4$-$R_X$-Ising$^4$-$R_Y$-Toffoli$^4$ architecture that empirically works very well, as illustrated in Figure \ref{fig:qnnencoder}.
In the quantum spectral encoder, customized for the 12-band Sentinel-2 data (but generally applicable to other multispectral modalities), we first divide the channel into three groups (each with 4 bands) to better capture local band correlations among adjacent channels \cite{wang2020super, zhang2022robust}, followed by independently learning neighboring channel relations to extract neighboring band correlations.
The quantum spectral encoder learns the local channel relations for all channel-wise sequences (cf., yellow $y_i$'s in Figure \ref{fig:qnnencoder}), meaning that all sequences share the same network weights in each group for achieving lightweight QEDNet, as graphically illustrated in Figure \ref{fig:qnnencoder}.
{The architectural differences between quantum spatial encoders and spectral encoders stem from their distinct functionalities.
The spatial encoder is designed to capture neighboring spatial information, whereas the spectral encoder focuses on spectral correlations.
Since capturing the spatial continuity in mangrove regions is relatively simple, we opt to adopt a shallower network for the spatial encoder.
On the contrary, considering the significance of spectral information for classification tasks, we developed a deeper quantum circuit with additional quantum gates for the spectral encoder.}

Afterward, inspired by the classical feature fusion concept \cite{song2018hyperspectral}, we create a quantum feature fusion block (FFB) to fuse the 12 spatial-spectral encoded feature maps.
As demonstrated in Figure \ref{fig:fusion}, the 12 channels are first classified into three neighboring channel groups, which are then utilized to obtain the local representative features with neighboring band correlation via QFM.
The QFM operates on its 3 input qubits, and is designed as the $R_Y$-Ising$^2$-$R_X$-Ising-$R_Y$-Toffoli$^3$ architecture that empirically works very well, as illustrated in Figure \ref{fig:fusion}.
Then, for each group, the remaining qubit (i.e., the 4th qubit) is directly merged with the output of the QFM, and this shortcut connection strategy is proposed to mitigate the barren plateau effect (a.k.a. quantum gradient vanishing effect) \cite{hyperqueen}.
%
%
The four neighboring channels in each group are then fused into one single channel, leading to the three output channels (corresponding to the three groups) that are fed into another QFM to obtain the globally fused channel, as illustrated in Figure \ref{fig:fusion}.
We remark that qubit 12 is copied twice, but this does not violate the no-cloning theorem \cite{NCT}, as the output of the spectral encoder has been measured as some statistics (cf. the last layer of the spectral encoder in Figure \ref{fig:qnnencoder}) that are not quantum states.

Another judicious strategy we adopted for achieving the lightweight network design is to increase the network expressibility when reducing the number of quantum neurons.
As the spectral property is more critical for classification tasks, we design the spectral encoder (and its subsequent spectral fusion module, i.e., QFM) to have the full expressibility (FE), meaning that the network can express all the valid quantum functions.
This property is stated in Theorem \ref{theorem: FE} below.
\begin{theorem} \label{theorem: FE}
The trainable quantum neurons deployed in the proposed quantum spectral encoder (cf. Figure \ref{fig:qnnencoder}) and QFM (cf. Figure \ref{fig:fusion}) can express any valid quantum unitary operator $U$, with some real-valued trainable network parameters $\{\beta_{k},\epsilon_{k},\eta_{k},\lambda_{k},\rho_{k}\}$ (for spectral encoder) and $\{\gamma_{k},\zeta_{k},\theta_{k},\phi_{k}\}$ (for QFM).
\hfill$\square$
\end{theorem}
The proof of Theorem \ref{theorem: FE} follows quite similar procedure to that of \cite[Theorem 2]{hyperqueen}, and is omitted here for conciseness.

Finally, the bicubic interpolation is adopted to upsample the spatial dimension of the QNN features before injecting them into the traditional CNN features, as shown in Figure \ref{fig:framework}.
To sum up, the proposed QNN branch $f_{\text{QNN}}$ is defined as
\begin{align}
f_{\text{QNN}}(\bT)&=\text{Bicubic}(\text{FFB}(\text{Encoder}_\text{Spe}(\text{Encoder}_\text{Spa}(\bT)))),
\end{align}
where the Bicubic$(\cdot)$ is the bicubic interpolation function; FFB$(\cdot)$ denotes the feature fusion block; $\text{Encoder}_\text{Spa}(\cdot)$ and $\text{Encoder}_\text{Spe}(\cdot)$ are quantum spatial and spectral encoders.
Finally, the QNN feature required in Equation \eqref{eq:twobranch} can be explicitly written as
\begin{align}
\bF_{\text{QNN}}&=f_{\text{QNN}}(\bX),
\end{align}
where $\bF_{\text{QNN}}\in \mathbb{R}^{H\times W }$ represents the QNN feature extracted from the QNN branch for determining the final classification result.
By following the aforementioned architecture, the lightweight QNN branch can be effectively implemented.

To conclude this section, we have created a lightweight dual-branch model combining the CNN and QNN tracks, as shown in Figure \ref{fig:framework}.
Due to their parallel cooperation, there is no need to create a very deep network with heavy parameters.
One can observe that the suggested CNN track is straightforward under the incorporation of valuable quantum features.
Furthermore, the QNN branch design is also explainable (with FE) and lightweight, making the proposed dual-branch framework a simple yet efficient model.
Additionally, this symmetric design highlights the equal significance of the CNN/QNN track, signifying the equal contributions of both CNN/QNN features.
Echoing the desire of acquiring new feature information, the QNN track (cf. Figure \ref{fig:qnnencoder} and Figure \ref{fig:fusion}) is adopted to help enhance classification accuracy, as will be experimentally proved in Section \ref{sec: experiment}.

\section{Experimental Results}\label{sec: experiment}
In this section, we will first introduce experimental settings, following demonstrate the effectiveness of the proposed QEDNet through both qualitative and quantitative assessments, and finally evaluate the efficacy of the quantum entangled feature (i.e., QNN track).
Section \ref{sec:experimentsetting} introduces the dataset sources and the network hyperparameters used for training QEDNet.
Section \ref{sec:Quantitative and Qualitative Results} presents the qualitative and quantitative assessment results across different countries.
In Section \ref{sec:ablation}, we conduct ablation studies to systematically assess the efficacy of the quantum entangled features (i.e., QNN track) in the dual-branch architecture.

\subsection{Experimental Settings}\label{sec:experimentsetting}

\begin{table}[t]

\captionsetup{justification=centering, labelsep=newline}
    \caption{Location, image size, and the usage of the Sentinel-2 data used in our experiments.}
    \label{table:Location Table}
    \centering
    
    \renewcommand{\arraystretch}{1.3}
    \scalebox{0.85}
    {
    \begin{threeparttable}
    \begin{tabular}{c|cccc}
        \hline\hline
        ID         & {Year}  &{(Lon,~\!Lat)*}    & {Image Size}        &{Usage}\\
         \hline\hline
        El Salvador-1  &    2018  & (-90.137,~\!13.643) & 1500$\times$2794 & {Training/Validation}\\
        
        El Salvador-2  &    2018  & (-89.342,~\!13.118) & 4501$\times$12420 & {Training/Validation}\\
        
        El Salvador-3  &    2018  & (-87.920,~\!13.320) & 2248$\times$2017 & {Training/Validation}\\
        
        Nicaragua-1    &    2018  & (-87.610,~\!12.713) & 4489$\times$6986 & {Training/Validation}\\
        
        Nicaragua-2   &    2018  &  (-87.401,~\!12.244) & 5228$\times$6210 & {Training/Validation}\\
        
        Nicaragua-3   &    2018  &   (-83.496,~\!14.397) & 6783$\times$4657 & {Training/Validation}\\
        
        Bangladesh-1   &    2018  &   (89.014,~\!21.673)   & 6345$\times$4116 &{Training/Validation}\\
        
        Bangladesh-2   &    2018  &   (89.493,~\!21.687)   & 6092$\times$4416 &{Training/Validation}\\
        
        Bangladesh-3   &    2018  &   (89.917,~\!21.769)   & 3815$\times$10519 &{Training/Validation}\\
        
        \hline

        Myanmar-1   &    2018  &  (98.346,~\!12.316) & 3045$\times$1980    & {Testing}\\
        
        Myanmar-2   &    2018  &  (97.347,~\!16.175) & 4265$\times$3807    & {Testing}\\

        Thailand-1         &    2018  &  (98.337,~\!9.411)  & 6045$\times$3778 & {Testing}\\
        
        Thailand-2         &    2018  &  (102.411,~\!12.058)  & 3022$\times$3022 & {Testing}\\     

        Cambodia-1     &    2018  &  (103.310,~\!10.956)   & 3788$\times$6058 & {Testing}\\
        
        Cambodia-2     &    2018  &  (102.874,~\!11.365)   & 5302$\times$3787 & {Testing}\\
        
        \hline\hline
        
        \end{tabular}

    \vspace{0.1cm}
     
    \begin{tablenotes}
		\item * denotes the location of the bottom-left corner pixel of the image.
    \end{tablenotes} 
   \end{threeparttable}}
\end{table}

\textit{1) Dataset Description:}
In this paper, 200 Sentinel-2 MSIs are employed for training the proposed QEDNet, including 180 training and 20 validation data.
Besides, all these Sentinel-2 datasets (i.e.,  Level-2A product\footnote{\scriptsize\url{https://sentinel.esa.int/documents/247904/685211/Sentinel-2_User_Handbook}.} with atmospheric correction) were acquired over 2018 and downloaded from the Google Earth Engine (GEE) \cite{GEE}.
Specifically, our study uses Aerosols, Blue, Green, Red, Red Edge 1, Red Edge 2, Red Edge 3, NIR, Red Edge 4, Water vapor, SWIR 1, and SWIR 2 bands for model design.
It is important to note that the Sentinel-2 images obtained from GEE have a lower resolution in certain bands and undergo preprocessing by the Sentinel-2 super-resolution software \cite{SSSS} to ensure consistent 10m resolution for all spectral bands.
{From a different perspective, Chen \textit{et al.} \cite{chen2024noval} collected additional data to train transformer-based super-resolution models \cite{liu2023ashformer,liu2024seismic} for enhancing the spatial resolution of Landsat data.}
Regarding the ground-truth map (i.e., pre-label) for training, we adopt the mangrove class label in ``Coastal Habitat Mapping: Mangrove and Pond Aquaculture Conversion" project from Clark Labs \cite{CLARK_LAB}.
Although mangroves and tides vary over time, it should be emphasized that we mainly focus on developing the MM algorithm using one single input (i.e., Sentinel-2 MSI data) without additional intertidal zone data collection.
{Table \ref{table:Location Table} summarizes the location (in terms of the longitude (Lon) and latitude (Lat) of the bottom-left corner pixel), retrieval year, and image size of Sentinel-2 data for the experimental investigation. 
Specifically, the study areas for training the model include El Salvador and Nicaragua in Central America, as well as Bangladesh in South Asia.}
For the assessments, the three testing sub-scenes (i.e., Myanmar, Thailand, and Cambodia data) were acquired on 26 Dec. 2018, 25 Dec. 2018, and 25 Dec. 2018, respectively.
%
%
\\

\textit{2) Network Hyperparameters and Computer Facilities:}
For training the proposed QEDNet, we adopt the AdamW optimizer \cite{loshchilov2018decoupled} with a learning rate decay strategy.
The initial learning rate is set as $1\times10^{-4}$ under the cosine annealing scheme. 
During the training process, the binary cross-entropy loss is employed.
The training process stops when the validation loss converges or reaches 200 epochs at maximum, indicating acceptable performance.
To evaluate the performance of the proposed model, we choose the checkpoint with the best kappa statistic ($\kappa$) in validation data for the following testing experiments.
In addition, the details of the computational setup and resources are outlined below.
All experiments are conducted on a desktop computer with an NVIDIA RTX 3090 GPU and an Intel Core i9-10900K CPU (3.70 GHz) with 128 GB of RAM, while the numerical computing environments for employing NN and index-based methods are Python 3.11.9 and Mathworks Matlab R2021b, respectively.
\subsection{Quantitative and Qualitative Analyses}\label{sec:Quantitative and Qualitative Results}
\begin{figure*}[t]
	\centerline{\includegraphics[width=1\textwidth]{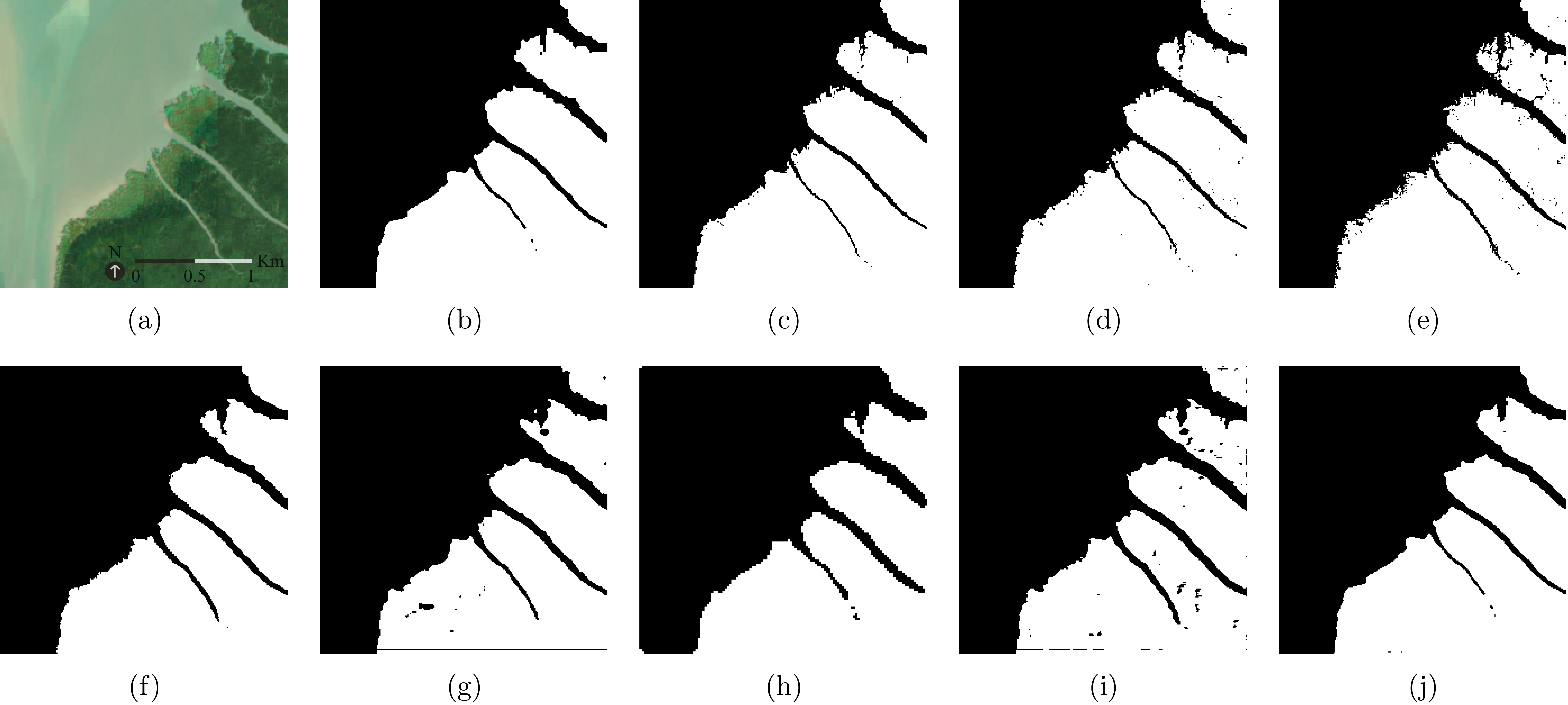}}
	\centering
	\caption{Qualitative study using Myanmar data. (a) RGB reference. (b) Ground-truth map. Classification maps obtained by (c) NDVI, (d) MMRI, (e) MVI, (f) GC-UNet, (g) DCNN, (h) ME-Net, (i) Capsules-UNet, and (j) the proposed QEDNet.}
	\label{fig:sim1}
\end{figure*}
\begin{figure*}[t]
	\centerline{\includegraphics[width=1\textwidth]{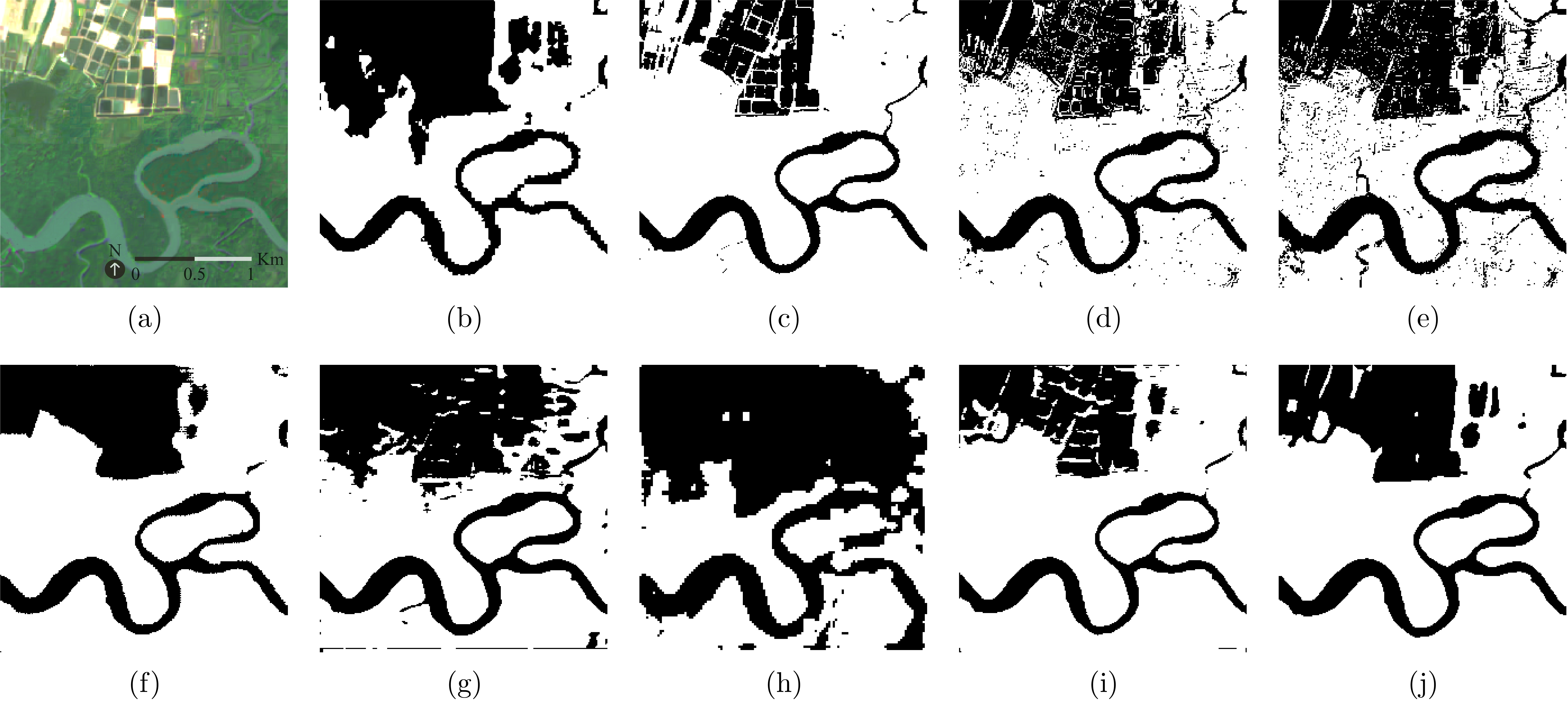}}
	\centering
	\caption{Qualitative study using Thailand data. (a) RGB reference. (b) Ground-truth map. Classification maps obtained by (c) NDVI, (d) MMRI, (e) MVI, (f) GC-UNet, (g) DCNN, (h) ME-Net, (i) Capsules-UNet, and (j) the proposed QEDNet.}
	\label{fig:sim2}
\end{figure*}
\begin{figure*}[t]
	\centerline{\includegraphics[width=1\textwidth]{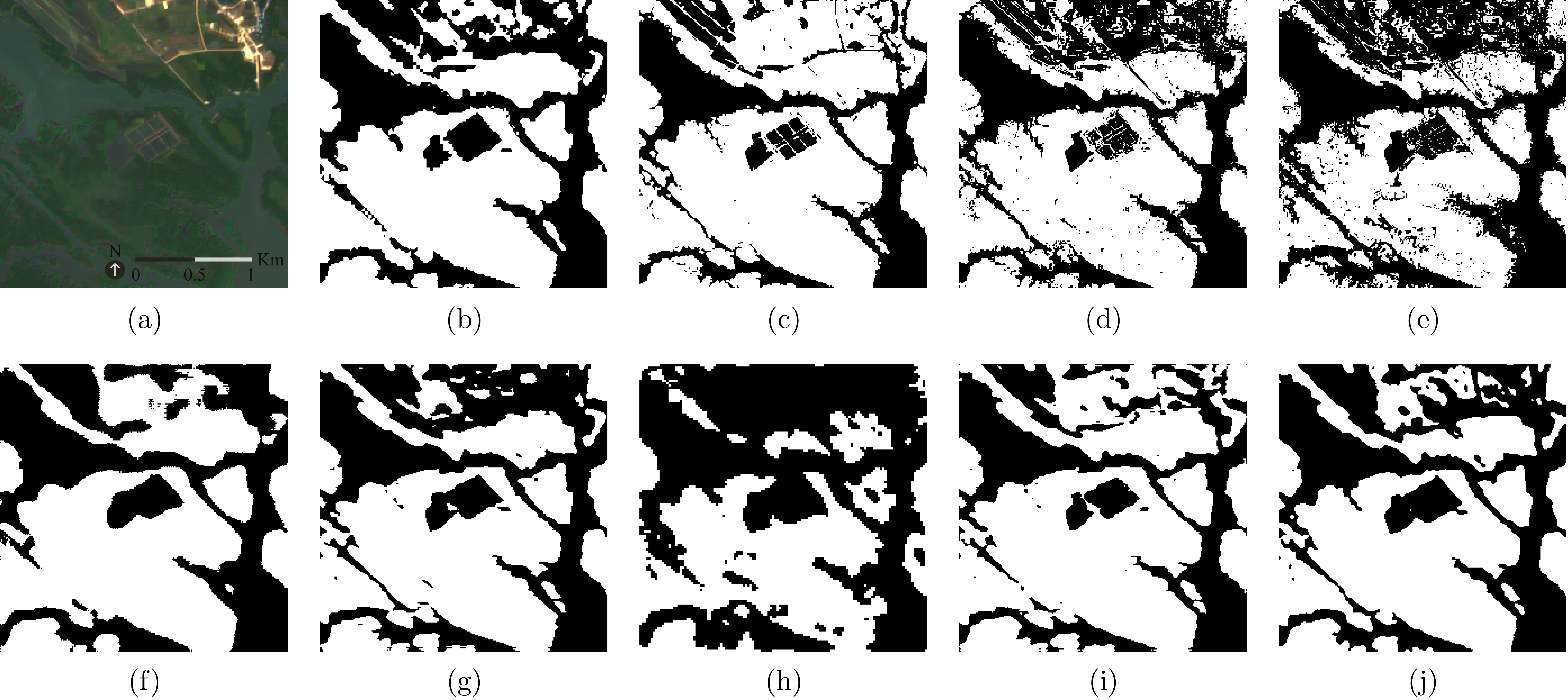}}
	\centering
	\caption{Qualitative study using Cambodia data. (a) RGB reference. (b) Ground-truth map. Classification maps obtained by (c) NDVI, (d) MMRI, (e) MVI, (f) GC-UNet, (g) DCNN, (h) ME-Net, (i) Capsules-UNet, and (j) the proposed QEDNet.}
	\label{fig:sim3}
\end{figure*}

In this section, we demonstrate the effectiveness of our proposed QEDNet in comparison to benchmark index-based and NN-based methods, including normalized difference vegetation index (NDVI) \cite{tucker1979red}, {modular} mangrove recognition index (MMRI) \cite{diniz2019brazilian}, mangrove vegetation index (MVI) \cite{MVI_paper}, classical UNet with global context extraction (GC-UNet) \cite{dong2021gc}, deep convolutional neural network (DCNN) \cite{iovan2020deep}, multiple features extracted deep network (ME-Net)\cite{MENetrs21}, and capsule mechanisms-aided UNet (Capsules-UNet)\cite{capsulenetrs21}. 
Besides, the threshold setting of index-based methods is summarized below. 
According to the default setting in \cite{adi2016detection}, we set the threshold value as $0.33$ for moderate mangrove density cases in our study.
As for MMRI and MVI, we adhere to the refined threshold established in recent literature \cite{codemm}.
The threshold values are defined as $-0.27$ and $2.6$, respectively, for enhancing their qualitative and quantitative results.
However, it should be noted that threshold-tuning in index-based methods requires additional time and effort, as considered a disadvantage compared to adaptive NN model-based methods.

The testing mangrove regions are mainly in Southeast Asia, which houses over one-third of the global mangrove forest and the highest diversity of mangrove species \cite{richards2016rates}.
As demonstrated in Figures \ref{fig:sim1}, \ref{fig:sim2}, and \ref{fig:sim3}, $256\times256\times12$ testing sub-scenes cover Myanmar, Thailand, and Cambodia regions.
Note that the testing data is sourced from different nations compared to the training/validation data. 
This data splitting approach helps us evaluate the methods' robustness, avoiding potential overfitting problems \cite{codemm}.
To evaluate the performance of the studied method, we assess them with various scenarios, from simple to hard ones, including coastline (cf. Figure \ref{fig:sim1}), riverine agricultural landscape (cf. Figure \ref{fig:sim2}), and irregular and complex terrain (cf. Figure \ref{fig:sim3}). 
The experimental results for the Myanmar data are presented in Table {\ref{table:simall}} and Figure \ref{fig:sim1}, while the results for the Thailand data are presented in Table {\ref{table:simall}} and Figure \ref{fig:sim2}. 
Additionally, the results of the Cambodia data are summarized in Table {\ref{table:simall}} and Figure \ref{fig:sim3}.
\begingroup
\setlength{\tabcolsep}{6pt} 
\renewcommand{\arraystretch}{1.2} 
\begin{table*}[t]
\begin{center}
\caption{Quantitative comparison on three real multispectral datasets.} 
\label{table:simall}
	\scalebox{1.2}{
\begin{tabular}{|cc|ccc|ccccc|}
\hline
& & \multicolumn{3}{c|}{Index-Based Approach}& \multicolumn{5}{c|}{Neural Network-Based Approach}\\
\hline
Dataset & Index & NDVI & MMRI & MVI & GC-UNet & DCNN & ME-Net & Capsules-UNet & QEDNet\\
\hline
\multirow{3}{*}{Myanmar} 
& OA ($\%$) & 97.97 & 97.84 & 96.83 & 97.33 & 96.19 & 95.36 & 95.87 & {\bf 98.95}\\
& AA ($\%$) & 97.98 & 97.81 & 96.59 & 97.05 & 95.78 & 94.87 & 95.42 & {\bf 98.87} \\
& $\kappa$ & 0.959 & 0.956 & 0.936 & 0.946 & 0.922 & 0.906 & 0.916 & {\bf 0.979}\\
\hline
\multirow{3}{*}{Thailand} 
& OA ($\%$) & 82.95 & 85.36 & 84.43 & 85.69 & 84.53 & 76.10 & 83.83 & {\bf 89.33}\\
& AA ($\%$) & 77.46 & 82.80 & 83.57 & 82.66 & 84.35 & 80.06 & 79.61 & {\bf 86.77}\\
& $\kappa$ & 0.600 & 0.676 & 0.667 & 0.680 & 0.674 & 0.539 & 0.630 & {\bf 0.762}\\
\hline
\multirow{3}{*}{Cambodia} 
& OA ($\%$) & 86.76 & 86.51 & 84.19 & 84.07 & 84.34 & 80.12 & 83.55 & {\bf 90.12}\\
& AA ($\%$) & 84.94 & 86.29 & 85.16 & 83.70 & 84.17 & 82.08 & 82.15 & {\bf 89.98}\\
& $\kappa$ & 0.721 & 0.724 & 0.685 & 0.674 & 0.681 & 0.612 & 0.656 & {\bf 0.798}\\
\hline

\end{tabular}}
\vspace{-0.3cm}
\end{center}

\end{table*}  
\endgroup
\begin{table*}[t]
\caption{Computational time (measured in seconds (sec.)) taken by various classification methods under investigation.
Among neural network-based methods that have taken the spatial continuity into consideration, the proposed QEDNet is the fastest due to its lightweight dual-branch design.
Though index-based methods are even faster, they ignore the continuity nature of MM, thus limiting their classification performance.}
	\begin{center}
 \setlength\tabcolsep{12pt}
	\renewcommand\arraystretch{1.76}
		\begin{tabular}{c|c|c|c|c|c|c|c|c} \hline\hline
        &\multicolumn{3}{c|}{Index-Based Approach} & \multicolumn{5}{c}{Neural Network-Based Approach} \\
        \hline
		 Method&	NDVI  &  MMRI  &  MVI 	&  GC-UNet  &  DCNN  &  ME-Net  &  Capsules-UNet  
		 &  QEDNet    \\
			\hline
		Time (sec.)  & {\bf 0.0006} & 0.0007 & 0.0008 & 0.8517 & 0.3188 & 0.6620 & 1.0257 & {\bf 0.1985}\\

			\hline\hline
		\end{tabular}
	\end{center}
	\label{table:time}
\end{table*}

For quantitative assessment, three commonly used objective metrics are adopted to assess the performance of the studied methods, including overall accuracy (OA) \cite{OA2}, average accuracy (AA) \cite{AA}, and Cohen's kappa coefficient ($\kappa$) \cite{kappa}.
The definitions of these metrics are detailed below.

The OA index assesses the percentage of correctly classified samples out of the total number of samples, defined as follows:
\begin{equation*} 
	\text{OA}= {\sum_{n=1}^{N} P_{nn}},
\end{equation*}
where $N=2$ represents the total number of classes, and $P_{nn}$ represents the probability of accurately predicting class $n$, while $P_{ij}$ denotes the probability of class $i$ being predicted when the actual class is $j$.

The AA index calculates the accuracy for each class independently and then computes the mean of these individual class accuracies, defined as follows:
\begin{equation*} 
	\text{AA} =\frac{1}{N}\sum_{n=1}^{N}\frac{P_{nn}}{P_{n.}},
\end{equation*}
where $P_{n.}=\sum_{n'=1}^N P_{nn'}$.
Since AA provides equal weight to each class, it is not dominated by a specific class, making it more accurate than OA, especially in scenarios with an imbalanced distribution.

The kappa statistic ($\kappa$) further considers the possibility of chance agreement ($p_{e}$) \cite{kappa}, which is defined as:
\begin{equation*}
\begin{aligned}
\kappa &= \frac{\text{OA}-p_{e}}{1-p_{e}}, \\
p_{e} &= \sum_{n=1}^N P_{n.}P_{.n},
\end{aligned}
\end{equation*}
\\
where $P_{.n} = \sum_{n'=1}^N P_{n'n}$.
In summary, the kappa coefficient provides a more precise performance evaluation by considering the accuracy results (OA) and the randomly correct results  ($p_e$).
If the performance of the testing model significantly surpasses random guessing, then the kappa statistic ($\kappa$) will be close to $1$.
As shown in Figures \ref{fig:sim1}, \ref{fig:sim2}, and \ref{fig:sim3}, we examine three sub-scenes across different countries to assess the effectiveness and universality of the studied method in quantitative evaluation.
Furthermore, the quantitative assessments of three testing sub-scenes are summarized in Table {\ref{table:simall}}.
The best performance is highlighted in bold font, indicating the highest OA/AA/$\kappa$ values.
Not surprisingly, the proposed QEDNet achieves state-of-the-art performance thanks to the quantum-entangled feature (i.e., QNN track).
To better understand the efficiency of the quantum unitary feature (i.e., QNN track), we conduct detailed ablation studies to evaluate the effectiveness of the QNN track in Section \ref{sec:ablation}.
Besides, benchmark methods do have some spatial continuity distortion and lower accurate values, as discussed below.

{To ensure a fair comparison and robust evaluation, training/validation datasets and testing datasets are constructed from completely non-overlapping nations, as shown in Table \ref{table:Location Table}.
This approach guarantees that the performance evaluation is performed on completely new data (i.e., different nations), thereby reflecting the method's generalizability.
Thus, we can observe that performance varies across different datasets due to their untrained nature and landscape complexity.}

First, we test the basic coastline landscape in the Myanmar region, as shown in Figure \ref{fig:sim1}.
Figure \ref{fig:sim1}(a) and \ref{fig:sim1}(b) represent the RGB reference and the ground-truth map for the Myanmar data, where white-colored pixels depict the mangrove region and black-colored pixels indicate the opposite.
Classification results from NDVI, MMRI, MVI, GC-UNet, DCNN, ME-Net, Capsules-UNet, and the proposed QEDNet are demonstrated from Figure \ref{fig:sim1}(c) to \ref{fig:sim1}(j), respectively.
In Figure \ref{fig:sim1}(e), \ref{fig:sim1}(g), and \ref{fig:sim1}(i), it is apparent that the MVI, DCNN, and Capsules-UNet classification results contain different levels of speckle noise, leading to the lower accuracy results in the third, fifth, and seventh {columns} of Table {\ref{table:simall}}.
Another complex scenario involving the river and farm landscape is being examined, as depicted in Figure \ref{fig:sim2}(a) and \ref{fig:sim2}(b).

The primary challenge is how to effectively differentiate the crops from the mangrove vegetation in the top areas.
For example, NDVI, MMRI, and Capsules-UNet fail to accurately classify the farm region, as shown in Figure \ref{fig:sim2}(c), \ref{fig:sim2}(d), and \ref{fig:sim2}(i).
In addition, ME-Net cannot capture the correct shape for spatial continuity in Figure \ref{fig:sim2}(h).
As illustrated in Figure \ref{fig:sim3}(a) and \ref{fig:sim3}(b), the Cambodia sub-scene features an irregularly shaped mangrove region as a more complex challenge.
Hence, it is suitable for studied methods to verify their effectiveness and resilience against disturbances.
Surprisingly, besides the proposed QEDNet, other methods do not adequately capture the texture and shape in the top area, as shown from Figure \ref{fig:sim3}(c) to \ref{fig:sim3}(i).
Not to mention that MMRI and MVI have some noisy distortion, while GC-UNet and ME-Net misclassify the river as mangrove in the lower-left region.
To conclude, across three testing scenarios, only the proposed QEDNet maintains high performance when the testing data gets harder.
As a comparison, the performance of benchmark methods declines significantly in the challenging Thailand and Cambodia data, unlike their performance in the simple Myanmar sub-scene.
\begin{table}[t]
\caption{Comparisons of model size and complexity. Note that M and G indicate $10^6$ and $10^9$. }
\renewcommand\arraystretch{1.2}
\begin{center}
\setlength{\tabcolsep}{1.6mm}
\scalebox{1.15}{
\begin{tabular}{c|c|c}
\hline
\hline

\multirow{1}{*}{Methods} & \multirow{1}{*}{\#Param.}  & \multirow{1}{*}{FLOPs}  

\\
\hline

\multirow{1}{*}{GC-UNet} 
& 1.1M & \bf\underline{2.96G} 

\\

\hline
\multirow{1}{*}{DCNN} 
& 1.43M & 103.88G 

\\

\hline
\multirow{1}{*}{ME-Net} 
& 141.9M & 18.07G 

\\

\hline
\multirow{1}{*}{Capsules-UNet} 
& \bf0.57M & 25.03G 

\\

\hline
\multirow{1}{*}{QEDNet} 
& \bf\underline{0.09M} & \bf7.59G 

\\
\hline
\hline
\end{tabular}}
\label{tab:ALL_FLOPS}
\end{center}
\end{table}
Finally, the computational time among studied methods is summarized in Table \ref{table:time}.
Notably, the proposed QEDNet is the fastest among NN model-based methods due to its lightweight design.
Though index-based methods are even faster, they ignore spatial continuity and lead to unsatisfactory results when the testing scene becomes harder.
On the other hand, the computational time does not consider the hidden costs, such as the effort and time to find the optimal threshold parameters for better performance.
From the user's perspective, all the methods studied are sufficiently fast for real-time applications.

{To sum up, we have tested various scenarios from easy to hard to evaluate the generalizability among studied methods.
From Table \ref{table:simall}, we can easily observe that the performance of peer methods declines when facing complicated terrains.
For NN-based methods, the decline primarily results from overfitting problems, indicating that the testing outcomes are not always as satisfactory as the training outcomes.
For index-based methods, threshold setting has a sensitive impact on the final results.
Therefore, if the testing scenario is too complicated to classify by the threshold, it will lead to unsatisfactory results.
On the contrary, the proposed QEDNet utilizes two different informative CNN and QNN features, thereby outperforming all studied methods.
It demonstrates the highest effectiveness for MM in coastal and riverine agricultural areas as well as irregular and complex terrain landscapes, as shown in Figures \ref{fig:sim1} to \ref{fig:sim3}.}

{To gain a better understanding of the lightweight characteristics of QEDNet, we have further analyzed the parameters and the floating point operations (FLOPs).
As shown in Table \ref{tab:ALL_FLOPS}, QEDNet has the fewest parameters and the second-fewest FLOPs compared to other benchmark NN-based methods.
Meanwhile, the lightweight-designed QEDNet demonstrates its robustness and universality under the OA/AA/$\kappa$ metrics across various testing scenarios.
This fact demonstrates that QEDNet is indeed a lightweight and effective model for solving the MM problem.}
\subsection{Ablation Study}
\label{sec:ablation}
\begin{table}[t]
\caption{Ablation study for demonstrating the contribution of the QNN track in QEDNet, and for showing the irreplaceable role of the QNN features.}
\renewcommand\arraystretch{1.2}
\begin{center}
\setlength{\tabcolsep}{1.6mm}
\scalebox{1.15}{
\begin{tabular}{c c c |c c c }
\hline
\hline
\makecell[c]{Track 1 \\ (CNN)} & \makecell[c]{Track 2 \\(CNN)} & \makecell[c]{Track 2 \\ (QNN)}  & OA ($\%$) & AA ($\%$)  &  $\kappa$

\\
\hline
\checkmark  & &  & 89.34 & 89.25 & 0.782        
\\
\hline
\checkmark  & \checkmark & & 88.78 & 88.37 & 0.770
 
\\ 
\hline
\checkmark& &\checkmark  & {\bf 90.12} & {\bf 89.98} & {\bf 0.798} 

\\

\hline
\hline
\end{tabular}
}
\label{tab:ablation study}
\end{center}
\end{table}
As demonstrated in Sections \ref{sec:Quantitative and Qualitative Results}, QEDNet has exhibited state-of-the-art performance in qualitative and quantitative analyses across various nations.
To further analyze the effectiveness of the proposed dual-branch framework, we have conducted related ablation studies, as summarized in Table \ref{tab:ablation study} and introduced below.

In our design principles, the proposed dual-branch network combines fundamentally different CNN/QNN features to enhance classification performance.
Thus, we compare the performance of the CNN-QNN dual-branch and single-branch CNN models to evaluate the efficacy of the newly added QNN track.
Through the first and last rows of Table \ref{tab:ablation study}, one can see that combining the QNN feature does improve all evaluation metrics with higher accuracy.
However, one may question whether the performance improvement is solely due to the increasing parameter numbers but not the quantum entangled feature itself.
Therefore, we replace the QNN track with the CNN track in the dual-branch architecture for further investigation, thereby evaluating the respective performance of the CNN-QNN and CNN-CNN dual-branch architectures.
Note that the CNN-CNN dual-branch model does not share weights in the respective CNN tracks (cf. track1 and track 2 in Table \ref{tab:ablation study}) for a fair comparison.
As evident from the second and last rows of Table \ref{tab:ablation study}, the dual-branch CNN-QNN model shows state-of-the-art performance, demonstrating significant improvement across all metrics.
Surprisingly, the performance of the dual-branch CNN-CNN model declines compared to the single-branch CNN model, indicating that the improved performance in the dual-branch CNN-QNN model is not due to increasing parameter.
Hence, this finding again shows the effectiveness of the QNN track as a side-proof, indicating the efficacy of the quantum entangled feature (i.e., QNN track).
To sum up, the quantum entangled feature of the QNN track indeed enhances the classification accuracy, as shown by the improvement of OA, AA, and $\kappa$ metrics in Table \ref{tab:ablation study}.

\section{Conclusions and Future Works}\label{sec: conclusion}

{This study presents QEDNet, a quantum-empowered deep learning framework that integrates the QNN with the CNN (cf. Figure \ref{fig:framework}) to further enhance the classification performance.
It is designed to address the challenging MM task, which maps mangrove forest distributions that are essential for climate change in the SDGs.
Remarkably, QEDNet is both parameter-tuning-free and very lightweight, achieving an average improvement of 7.36\% in $\kappa$ over state-of-the-art MM methods. 
Additionally, it surpasses NN-based approaches in computational speed, demonstrating its superior computing efficiency and practical applicability.
Ablation studies have proven that QNN features do provide radically new information for better decision-making results (cf. Table\ref{tab:ablation study}).
Intriguingly, when the QNN track is replaced by CNN with the same architecture, the resultant dual-CNN does not yield better performance even if it has an increasing number of network parameters.
These findings emphasize again the value of combining affine-computing CNN features with unitary-computing QNN features, which provide fundamentally different yet complementary information.
Therefore, we are able to conclude that the new QNN feature information contributes to complicated classification tasks and beyond, underscoring their importance in the proposed framework.
In the future, we plan to explore cutting-edge remote sensing applications by fusing the unitary-computing QNN features with traditional CNN features, and propose customized fusion network architectures for different target missions.}

\renewcommand{\thesubsection}{\Alph{subsection}}
\bibliography{ref}

\begin{IEEEbiography}[{\resizebox{0.9in}{!}{\includegraphics[width=1in,height=1.25in,clip,keepaspectratio]{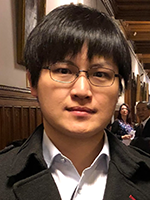}}}]
{\bf Chia-Hsiang Lin}
(S'10-M'18-SM'24)
received the B.S. degree in electrical engineering and the Ph.D. degree in communications engineering from National Tsing Hua University (NTHU), Taiwan, in 2010 and 2016, respectively.
From 2015 to 2016, he was a Visiting Student of Virginia Tech,
Arlington, VA, USA.

He is currently an Associate Professor with the Department of Electrical Engineering, and also with 
the Miin Wu School of Computing,
National Cheng Kung University (NCKU), Taiwan.
Before joining NCKU, he held research positions with The Chinese University of Hong Kong, HK (2014 and 2017), 
NTHU (2016-2017), 
and the University of Lisbon (ULisboa), Lisbon, Portugal (2017-2018).
He was an Assistant Professor with the Center for Space and Remote Sensing Research, National Central University, Taiwan, in 2018, and a Visiting Professor with ULisboa, in 2019.
His research interests include network science, 
quantum computing,
convex geometry and optimization, blind signal processing, and imaging science.

Dr. Lin received the Emerging Young Scholar Award from National Science and Technology Council (NSTC), in 2023,
the Future Technology Award from NSTC, in 2022,
the Outstanding Youth Electrical Engineer Award from The Chinese Institute of Electrical Engineering (CIEE), in 2022,
the Best Young Professional Member Award from IEEE Tainan Section, in 2021,
the Prize Paper Award from IEEE Geoscience and Remote Sensing Society (GRS-S), in 2020, 
the Top Performance Award from Social Media Prediction Challenge at ACM Multimedia, in 2020,
and The 3rd Place from AIM Real World Super-Resolution Challenge at IEEE International Conference on Computer Vision (ICCV), in 2019. 
He received the Ministry of Science and Technology (MOST) Young Scholar Fellowship, together with the EINSTEIN Grant Award, from 2018 to 2023.
In 2016, he was a recipient of the Outstanding Doctoral Dissertation Award from the Chinese Image Processing and Pattern Recognition Society and the Best Doctoral Dissertation Award from the IEEE GRS-S.
\end{IEEEbiography}

\begin{IEEEbiography}[{\resizebox{1in}{!}{\includegraphics[width=1in,height=1.25in,clip,keepaspectratio]{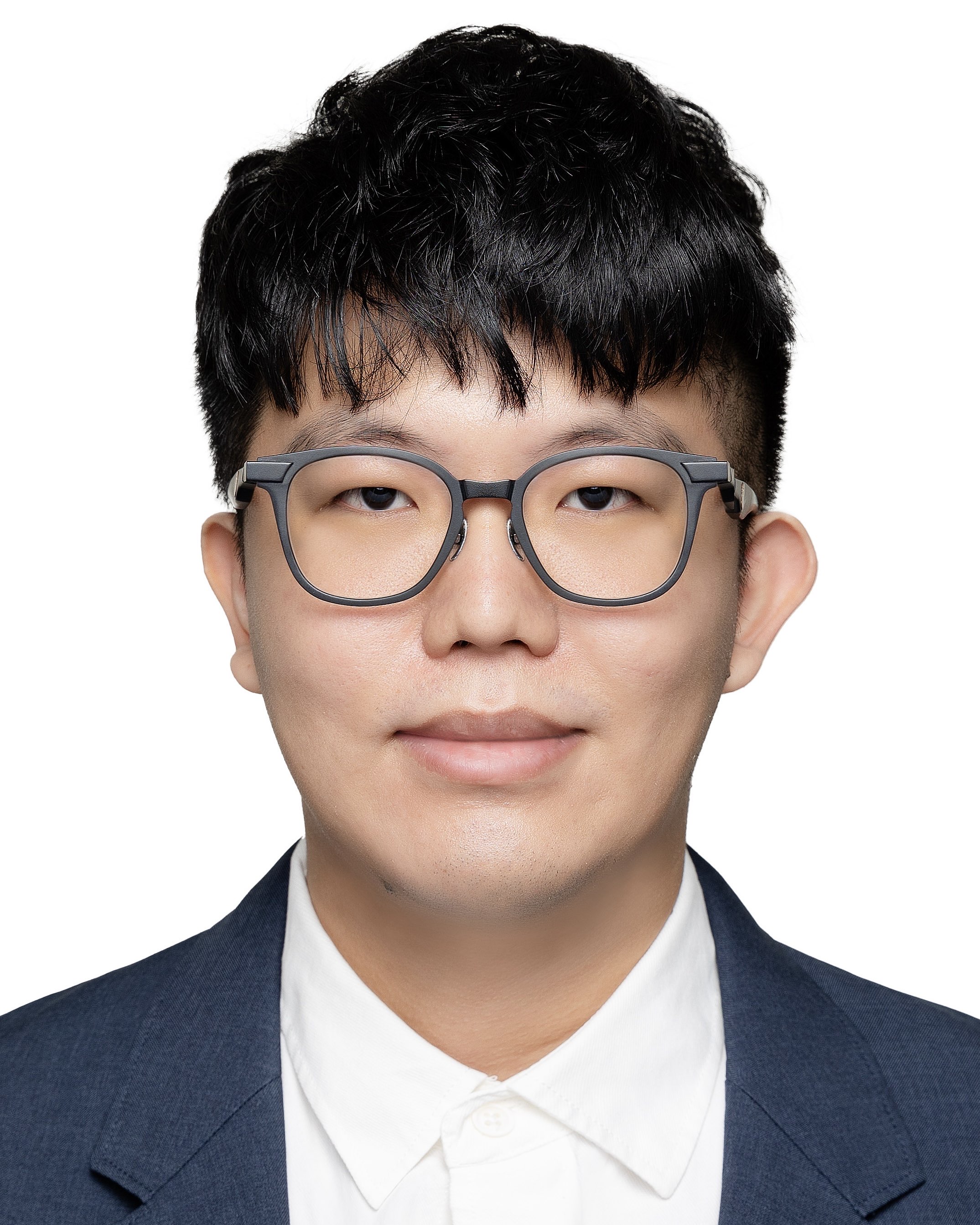}}}]
{\bf Po-Wei Tang}
(S'20)
received the B.S. degree from the Department of Electronic Engineering, National Changhua University of Education, Changhua, Taiwan, in 2018.
He is currently pursuing the Ph.D. degree with the Intelligent Hyperspectral Computing Laboratory, Institute of Computer and Communication Engineering, National Cheng Kung University (NCKU), Tainan, Taiwan.
His research interests include deep learning, convex optimization, tensor completion, and hyperspectral imaging.

Mr. Tang has received a highly competitive scholarship from NCKU, as well as the Pan Wen Yuan Award from the Industrial Technology Research Institute (ITRI) of Taiwan.
He has been selected as a recipient for the Ph.D. Students Study Abroad Program from the National Science and Technology Council (NSTC), in 2024.
	\end{IEEEbiography}

\begin{IEEEbiography}[{\resizebox{0.98in}{!}{\includegraphics[width=1in,height=1.25in,clip,keepaspectratio]{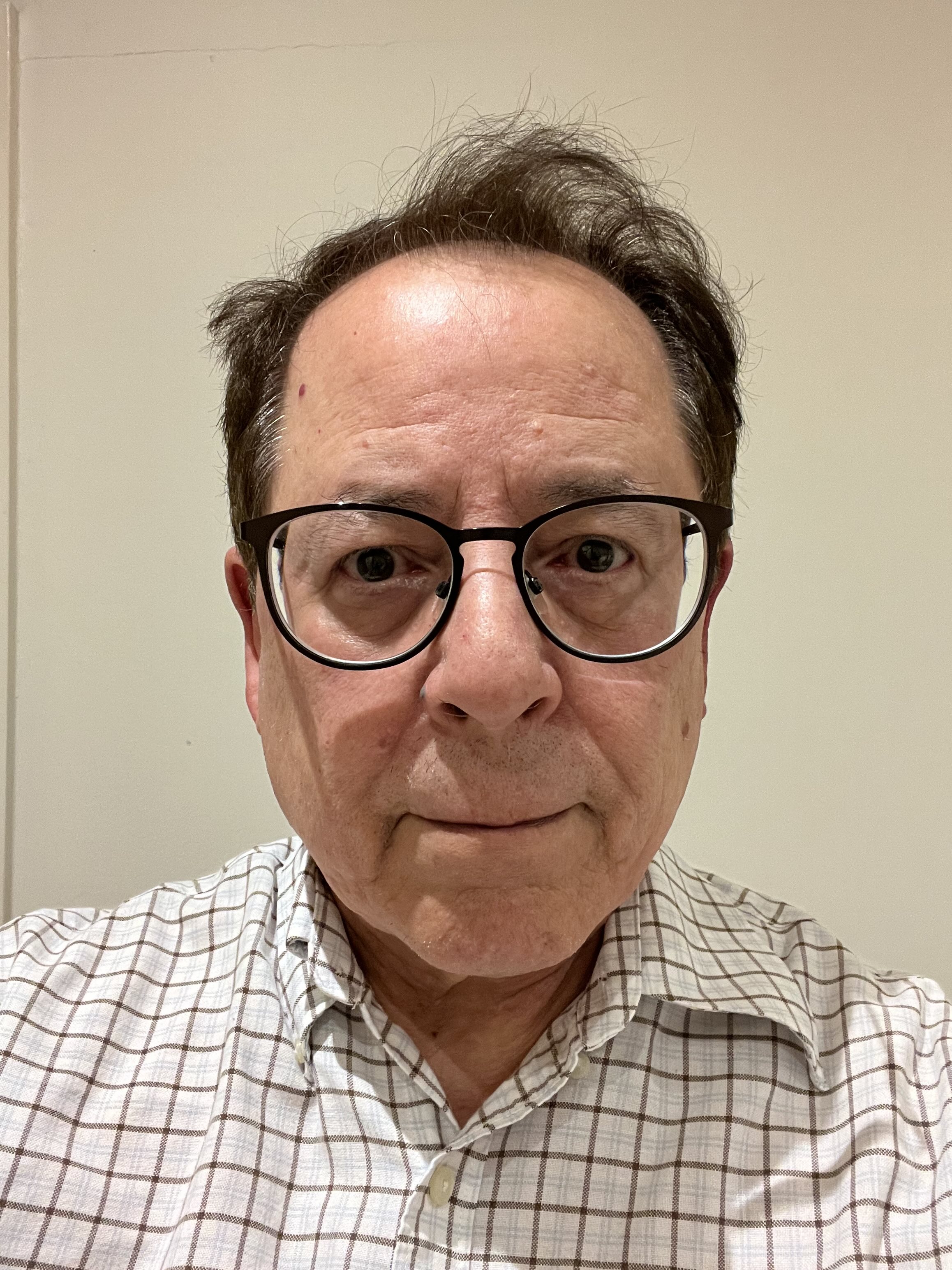}}}]
{\bf Alfredo R. Huete} received the M.Sc. degree in soil and plant biology from the University of California at Berkeley, Berkeley, CA, USA, in 1982, and the Ph.D. degree in soil and water science from The University of Arizona, Tucson, AZ, USA, in 1984. From 1984 to 2009, he was an Assistant Professor and an Associate Professor with The University of Arizona. He has over 20 years of experience in working on satellite mission teams, including the NASA-EOS MODIS Science Team, the New Millennium EO-1 Hyperion Team, and JAXA GLI Team, through which he developed the soil-adjusted vegetation index (SAVI) and the enhanced vegetation index (EVI). He is currently a Distinguished Professor with the University of Technology Sydney, Sydney, NSW, Australia, where he leads the Ecosystem Dynamics Health and Resilience Research Program in the Faculty of Science and is a Core Member of the Centre for Advanced Modelling and Geospatial Information Systems, Faculty of Engineering and Information Technology. He also leads the development of vegetation products for the Australian Terrestrial Ecosystem Research Network (TERN).
\end{IEEEbiography}

\end{document}